# Altmetrics: new indicators for scientific communication in Web 2.0

**Daniel Torres-Salinas** is a Research Management Specialist in the Evaluation of Science and Scientific Communication Group in the Centre of Applied Medical Research of the University of Navarra (Spain)
torressalinas@gmail.com

**Álvaro Cabezas-Clavijo** is a Contracted Researcher in the Faculty of Communication and Documentation of the University of Granada (Spain)
acabezasclavijo@gmail.com

**Evaristo Jiménez-Contreras** is a University Professor and Director of the Evaluation of Science and Scientific Communication Group in the Faculty of Communication and Documentation of the University of Granada (Spain)
evaristo@ugr.es

**ABSTRACT**

*This paper presents a review of altmetrics or alternative metrics. This concept is defined as the creation and study of new indicators for analysing scientific and academic research activity based on Web 2.0. The underlying premise is that variables such as mentions in blogs, number of tweets or saves of an article by researchers in reference management systems, may be a valid measure of the use and impact of scientific publications. In this respect, these measures are becoming particularly relevant, being at the centre of debate within the bibliometric community. Firstly, an explanation is given of the main platforms and indicators for this type of measurement. Subsequently, a study is undertaken of a selection of papers from the field of communication, comparing the number of citations received with their 2.0 indicators. The results show that the most cited articles within recent years also have significantly higher altmetric indicators. Next follows a review of the principal empirical studies undertaken, centering on the correlations between bibliometric and alternative indicators. To conclude, the main limitations of altmetrics are highlighted, alongside a reflective consideration of the role altmetrics may play in capturing the impact of research in Web 2.0 platforms.*

Science, scientific communication, information, communication, Internet, social networks, quantitative methods, Social Web; Web 2.0.

**1. Introduction.**

Altmetrics is a very new term, and can be defined as the creation and study of new indicators for the analysis of academic activity based on Web 2.0 (Priem & al., 2010). The underlying premise is that, for example, mentions in blogs, number of re-tweets or saves of articles in reference management systems, may be a valid measure of the use of scientific publications. However, measuring the visibility of science on the Internet is not a new phenomenon. The origin of altmetrics arose in the nineties with webometrics, the quantative study of the characteristics of the web (Thelwall & al., 2005). This was derived from the application of bibliometric methods to online sites, and encompasses various disciplines including communication. Despite the web playing an increasingly important role in social and economic relations, this discipline has not been capable of overcoming certain limitations inherent in the methodologies, methods and information sources used. However, it has contributed a complementary perspective to the traditional analysis of citations by means of the study of links, mailing list communications or analysis of the structure of the academic web. Shortly afterwards, the consolidation of scientific communication by journals and electronic media such as repositories opened the door to new indicators.

The so-called «bibliometrics usage» (Kurtz & Bollen, 2010), based on downloads of scientific materials, reveals that indicators of use of publications measure a different dimension to that of bibliometric indicators (Bollen & al, 2009), and demonstrate different behaviour patterns to citation (Schloegl & Gorraiz, 2010). With a view to, measuring scientific impact, these indicators offer complementary information. Without doubt, the idea that traditional bibliometric measures and the sources on which they base their calculations are insufficient permeates throughout the scientific community. This leads to the emergence of new indicators, such as SJR (González-Pereira & al., 2010) or the Eigenfactor (Bergstrom, West & Wiseman 2008), which are based on the idea of Google's PageRank algorithm. There is a clear symbiotic relationship between web based and bibliometric methods. This move is motivated by the dissatisfaction of many scientists with bibliometric methods, in particular the highly criticised Impact Factor (Seglen, 1997; Rossner, Van Epps & Hill., 2007), exacerbated by the appearance of new databases such as Scopus and Google Scholar. This search engine's power and coverage, but also its normalisation problems, illustrate both the wealth of academic information on the web, and the difficulty of adequately understanding and analysing this information (Torres-Salinas, Ruíz-Pérez & Delgado, 2009; Delgado & Cabezas-Clavijo, 2012).

It is in this context, with the arrival of Web 2.0 and scientists' gradual use of said platforms as tools for the diffusion and receipt of scientific information (Cabezas-Clavijo, Torres-Salinas & Delgado, 2009) and with part of the scientific community relatively receptive, that scientometrics 2.0 (Priem & Hemminger, 2010), or altmetrics (Priem & al., 2010), began to be discussed. Although, in a wider sense, any unconventional measure for the evaluation of science can be considered an alternative indicator, *sensu stricto* it would be more accurate to speak of indicators derived from 2.0 tools; that is to say, measures generated from the interactions of social web users (primarily but not exclusively scientists) with researcher produced material. One of the principal strengths of altmetrics lies in its provision of information at article level (Neylon & Wu, 2009), which enables assessment of the impact of papers beyond the bounds of publication sources. Various studies have stated that altmetrics can be used for measuring other levels of aggregation, such as journals (Nielsen, 2007) or universities (Orduña & Ontalba, 2012). Additionally, altmetrics offer a new perspective, considering the almost real time information provided on research impact. This monitoring, in the form of revision by peer collectives or peer revision following publication (Mandavilli, 2011), is undoubtedly an element that introduces new forms of scrutiny by the scientific community.

Taking into account the impact of Web 2.0 and its now central position within communication research, this paper undertakes a review of altmetrics, focusing on quantative studies of the same. Firstly, an explanation is given of the main platforms and indicators, followed by the comparative evaluation of a selection of communication papers showing the number of citations received and their 2.0 indicators. Next, a review of the principal empirical studies is undertaken, centering on the correlations between bibliometric and alternative indicators. To conclude, the main limitations of altmetrics are highlighted alongside a reflective consideration of the role altmetrics may play when it comes to understanding the impact of research in Web 2.0 platforms.

**2. Principal platforms and altmetric indicators.**

The placing on-line of bibliographic reference management systems and favourites, where personal libraries and researchers' references are regularly managed, has generated a series of original indicators. For example, the number of times a study has been marked as favourite (bookmarking) or the number of times it has been added to a bibliographic collection. Such indicators point to the reader interest aroused by scientific papers and the use made of them (Haustein & Siebenlist, 2011). On the other hand, some authors such as Taraborelli (2008), note that these indicators represent a form of quick review, by reflecting the degree to which papers are accepted by the scientific community. Among the most usual platforms for extracting these types of indicators are CiteUlike, Connotea or Mendeley (Li, Thelwall & Giustini, 2011). Of these, Mendeley currently arouses the most interest. According to its web page statistics, more than 2 millon users have uploaded a total of 350 millon documents, figures that mean an article's number of Mendeley readers has become one of the most accepted metrics for evaluating an articles impact within altmetrics.

**Table 1. Principal measurements proposed by altmetrics, classified according to type of platform, indicator and social network or platform**

| Type of platform | Type of indicator | Social network or platform | Examples of indicators |
|---|---|---|---|
| DIGITAL LIBRARIES AND REFERENCE MANAGEMENT SYSTEMS | Social bookmarking and digital libraries | General<br>● Delicious | ● Nº of times marked as favourite<br>● Nº of groups<br>● Nº of groups added to |
| | | Academic<br>● Citeulike<br>● Connotea<br>● Mendeley | |
| SOCIAL NETWORKS AND MEDIA | Mentions In social networks | General<br>● Facebook<br>● Google+<br>● Twitter | ● Number of likes<br>● Number of clicks<br>● Number of comments<br>● Number of times shared<br>● Numbern of mentions in tweets<br>● Number of retweets<br>● Retweets of leading users<br>● Etc. |
| | | Academic<br>● Academia.edu<br>● Research Gate | |
| | Mentions in blogs | General<br>● Blogger<br>● Wordpress | ● Number of blog citations<br>● Comments on the entry in blogs<br>● Systems of rating the entry |
| | | Academic<br>● Nature Blogs<br>● Postgenomic blog<br>● Research Blogging | |
| | Mentions in encyclopedias | ● Wikipedia<br>● Scholarpedia | ● Citations in the encyclopedia's entry |
| | Mentions in news promotion systems | General<br>● Reddit<br>● Menéame | ● Number of times on the title page<br>● Number of Clicks (moves)<br>● Number of comments on the news<br>● Punctuation of experts |
| | | Academic<br>● Faculty of 1000 | |

Other usual measurements are the mentions papers can receive in the multiple social networks in existence, these being a reflection of the diffusion and dissemination of publications (Torres-Salinas & Delgado, 2009). Normally, general social networks are used to calculate indicators, as in the case of Facebook or Twitter, by analysing the number of 'likes', the number of times an article is shared or the tweets and retweets received. Alternative metrics also include the blog citations received by scientific articles, especially in scientific blogs such as those included in the Nature Blogs or Research Blogging networks (Fausto & al., 2012). This is also true for the citations received by articles, journals and authors in the popular Wikipedia (Nielsen, 2007). These measurements are quantative approximations of the measure of interest aroused within the scientific community, and also amongst a general public, which transcend or compliment the impact of traditional citation indices. Finally, it is worth mentioning that news promotion systems such as Menéame or Reddit, or platforms with subject specialisation such as Documenea, can also offer indicators of research impact amongst a non-specialised public (Torres-Salinas & Guallar, 2009).

As can be seen in table 1, there exist a large number of indicators of distinct nature, origin and degree of normalisation. This means that the first difficulty faced when compiling information for a specific publication, and the subsequent altmetric calculation, is the high cost in time and effort. To solve this problem, a series of tools have emerged to help monitor impact. Generally, these types of platforms, once one or more documents are included, use a unique identification number such as the DOI or the PUBMEID to return the grouped metrics. Some of these tools are almetric.com, Plum Analytics, Science Card, Citedin or Impact Story. For scientific papers, statistics are normally presented from Facebook (Clicks, Shares, Likes or Comments), Mendeley (Readers, Number of Groups), Delicious, Connotea and Citeulike (Bookmarks) and Twitter (Tweets and Influential Tweets). In their favour, it has to be said that these tools enable the easy recuperation of statistics of collections of papers. However, they are limited by the presentation of contradictory results and only partially recover the statistics.

### 3. Altmetrics versus bibliometrics: examples in the field of communication.

In order to illustrate the tools and their derived indicators, data has been compiled from the 30 journal papers from the field of communication most cited in Web of Science for the years 2010, 2011 and 2012 (the ten most cited for each year). This sample has been compared with a random control group of another 30 papers, comprised of uncited articles from the same journals and years. In this way, the objective is to verify if a connection exists between the most cited articles and those that show superior data from alternative indicators. Once both samples of articles were downloaded from Web of Science (n=60; date: 04/02/2013), the altmetrics information was compiled using ImpactStory and Altmetric.com as sources. The following indicators were calculated for each article: mentions of the paper on Twitter, readers who have saved it in Mendeley and number of times it has been marked as favourite in Citeulike (table 2). The high occurrence of zeros among the most cited articles can be confirmed, in particular with regard to the indicators of Citeulike. This demonstrates one of the limitations of these statistics, as does the scant representation of some of these tools for reflecting scientific activitity.

**Table 2. Example of the number of citations and different altmetrics calculated for the ten most cited studies of 2012 in communication in Web of Science**

| Title of the article and of the studies | CITATIONS | TWEETS (TWITTER) | | READERS (MENDELEY) | | FAVOURITES (CITEULIKE) | |
|---|---|---|---|---|---|---|---|
| | WoS | IS | ALT | IS | ALT | IS | ALT |
| Epistemics in Action: Action Formation and Territories of Knowledge. Research on Language and Social Interaction | **13** | 0 | 0 | 20 | 20 | 0 | 0 |
| The Epistemic Engine: Sequence Organization and Territories of Knowledge. Research on Language and Social Interaction | **9** | 0 | 0 | 13 | 13 | 0 | 0 |
| Normalizing Twitter Journalism Practice in an Emerging Communication Space. Journalism Studies | **8** | 21 | 26 | 0 | 17 | 0 | 0 |
| Politics as Usual? Revolution, Normalization and a New Agenda for Online Deliberation. New Media & Society | **4** | 2 | 9 | 27 | 21 | 1 | 1 |
| The Dynamics of Audience Fragmentation: Public Attention in an Age of Digital Media. Journal of Communication | **4** | 0 | 5 | 0 | 33 | 0 | 0 |
| Pursuing a Response by Repairing an Indexical Reference. Research on Language and Social Interaction | **4** | 0 | 0 | 0 | 0 | 0 | 0 |
| It's a Dirichlet World: Modeling Individuals' Loyalties Reveals How Brands Compete, Grow, and Decline. Journal of Advertising Research | **3** | 0 | 0 | 0 | 0 | 0 | 0 |
| In 25 Years, Across 50 Categories, User Profiles for Directly Competing Brands Seldom Differ Affirming... Journal of Advertising Research | **3** | 0 | 0 | 0 | 0 | 0 | 0 |
| The Influence of Morality Subcultures on the Acceptance and Appeal of Violence. Journal of Communication | **3** | 0 | 4 | 5 | 0 | 1 | 1 |
| Grammatical Flexibility as a Resource in Explicating Referents. Research on Language & Social Interaction | **3** | 0 | 0 | 0 | 0 | 0 | 0 |

WoS = Web of Science; IS = Impact Story; ALT = Altmetric.com

The frequently cited articles were tweeted on more occasions than studies from the control sample (table 3). According to the first source (Impact Story), the cited articles were tweeted on average once more than the control sample, which did not receive any tweets. These figure increase to 2.5 and to 0.8 respectively, according to Altmetric.com. Although, due to the large number of papers not tweeted, the median in all cases is zero. Turning to Citeulike, the social bookmarking tool for scientists, the articles most cited between 2010 and 2012 were saved an average of 1.5 times (1.3 according to Altmetric.com), against 0.1 for the control sample; although only between 23% and 30% of the studies show values different to zero. However, the most representative data is that from Mendeley, where the most cited studies have been saved by an average of 18.6 readers (15.2 according to Altmetric.com), whilst the control sample shows an average of 4.6 readers (2.4 according to Altmetric.com). That is, the most cited papers are also saved more times by academics than uncited papers from the same journals. This indicator is the most representative of the amount by which between 57% and 62% of the articles, depending on the source consulted, present indicators different to zero.

**Table 3. Average, standard deviation and median of the altmetrics for a sample of 60 communication articles indexed in the Web of Science**

| | CITATIONS | TWEETS (TWITTER) | | READERS (MENDELEY) | | FAVOURITES (CITEULIKE) | |
|---|---|---|---|---|---|---|---|
| SAMPLE | WoS* | IS* | ALT | IS | ALT* | IS* | ALT* |
| CITED | 11.3 ± 6.1 (9.5) | 1.0 ± 3.9 (0) | 2.5 ± 6.1 (0) | 18.6 ± 25.7 (5.5) | 15.2 ± 19.1 (10) | 1.5 ± 3.4 (0) | 1.3 ± 3.4 (0) |
| NOT CITED | 0.0 ± 0.0 (0) | 0.0 ± 0.0 (0) | 0.8 ± 2.9 (0) | 4.6 ± 6.2 (2.5) | 2.4 ± 3.8 (0.5) | 0.1 ± 0.4 (0) | 0.1 ± 0.4 (0) |

*Statistically significant differences. Mann-Whitney Test; CI: 95%; p<0.05. Data expressed as Average ± Standard deviation (median). WoS: Web of Science; IS: Impact Story; ALT: Altmetric.com.

**4. Relationships between bibliometric indicators and altmetrics.**

An interesting underlying theme, in view of the data presented and the different studies that have been undertaken, is the relationship that exists between classic bibliometric indicators and the new metrics. These studies are of interest because they reveal whether the altmetrics correlate with papers' citations or if the opposite situation is produced, that is to say they reflect a new impact dimension. Clearly, in the sample of 60 communication studies, the correlation coefficients between citation in Web of Science and the altmetrics is low and of little significance (table 4). The highest achieved is between Pearson's correlation coefficient between citations and the number of readers of Mendeley, but it barely reaches 0.52.

**Table 4. Examples of studies of correlations between bibliometric indicators and altmetrics**

| Study | Sample used | Indicators compared | Correlations |
|---|---|---|---|
| **Results for the sample used in this study** | | | |
| Data presented in this study | 60 communication articles (see table 2 and table 3) | Citations Web of Science - Tweets | 0.09 Pearson<br>0.08 Spearman |
| | | Citations Web of Science - Mendeley | 0.52 Pearson<br>0.44 Spearman |
| | | Citations Web of Science - CiteUlike | 0.30 Pearson<br>0.46 Spearman |
| *Statistically significant differences. Mann-Whitney Test; p<0.01. Data calculated using Altmetric.com.* | | | |
| **Studies relating to Bibliometric Indicators and Altmetrics** | | | |
| Cabezas-Clavijo & Torres-Salinas 2010 | 8.945 published in the journal PLoS One | Citations Scopus - Nº Scientific Blog Links | 0.14 Pearson |
| | | Citations Scopus – Article Comments | 0.21 Pearson |
| Eysenbanch 2011 | 55 highly cited articles from the JMIR | Citations Google Scholar - Nº Tweets | 0.69 Pearson<br>0.36 Spearman |
| | | Citations Scopus - Nº Tweets | 0.54 Pearson<br>0.22 Spearman |
| Li, Thelwall, & Giuistini 2012 | 1,613 articles from Nature and Science published in 2007 | Citations Web of Science - Mendeley Bookmarks | 0.55 Pearson |
| | | Citations Google Scholar - Mendeley Bookmarks | 0.60 Pearson |
| | | Citations Web of Science - Citeulike | 0.34 Pearson |
| | | Citations Google Scholar - Citeulike | 0.39 Pearson |
| Bar-Ilan & al. 2012 | 1,136 articles by bibliometric researchers | Citations Scopus – Mendeley Bookmarks | 0.45 Spearman |
| | | Citations Scopus – Citeulike | 0.23 Spearman |
| Shuai, Pepe & | 70 articles deposited in the | Citations Google Scholar - Twitter mentions | 0.45 Pearson |

| Bollen 2012 | repository ARXIV | Downloads from arXiv - Twitter mentions | 0.55 Pearson |
| --- | --- | --- | --- |
| **Fausto & al. 2012** | 26,154 papers in 3,350 and reviewed in Researchblogging | Impact Factor – Blog views | 0.3 – 0.4 Pearson |
| | | Impact Factor – Blog Citations | 0.2 – 0.3 Pearson |
| | | Eigen Factor – Blog views | 0.3 – 0.4 Pearson |
| | | Eigen Factor – Blog Citations | 0.2 – 0.3 Pearson |

These results are in accordance with those obtained in other scientific papers (table 4). Cabezas-Clavijo & Torres-Salinas (2010) demonstrate that, for articles published in the journal PloS One, there is no connection between citation and comments and blog links received. A similar situation occurs if the Impact Factor or the EigenScore are used instead of citations (Fausto, 2012). With regard to the correlation between citation and Twitter, Eysenbanch (2011) observes very poor correlations in a global sample of 286 articles. The highest correlations between bibliometric indicators and altmetrics are produced, above all, when the former are compared with the number of readers in Mendeley; this is demonstrated by Li, Thelwall & Giuistini (2011) using the citations received in Google Scholar as an indicator. The correlation with Mendeley reaches 0.60 for a collection of papers published in «Science» and «Nature». If more specific fields of knowledge such as bibliometrics are taken into account, the correlation between readers in Mendeley and citations in Scopus rises to 0.45 (Bar-Ilan & al., 2012), a figure similar to that arrived at in this paper.

Therefore, in scientific literature to date, the correlation between any of the altmetrics and the number of citations remains to be convincingly demonstrated. However, evidence does exist of a certain association between highly cited or frequently downloaded and highly tweeted articles. For example Eysenbanch (2011), on isolating 55 highly cited articles from his sample, showed that in 75% of cases they were also highly tweeted, reaching a correlation coefficient of 0.69, the highest calculated to date. In addition, Shuai, Pepe & Bollen (2012), working with a sample of pre-prints deposited in ArXiv, observed greater download levels for papers promptly disseminated on Twitter. In the present case the most cited sample (table 3) also had higher rates of activity in social networks.

The results presented in table 4 suggest that altmetrics measure a dimension of scientific impact that is still to be determined. As stated by Priem, Piwowar & Hemminger (2012), there is a need for additional research into the validity and precise significance of these metrics, as, for example, in the case of the readers of Mendeley (Bar-Ilan, 2012). It seems apparent that altmetrics capture a different dimension, which could be entirely complementary to citation, given that the different platforms have audiences more diverse than the merely academic. If, for example, the phenomenon is observed from the other perspective, that of papers with greater altmetric impact, the studies most widely diffused across social networks in 2012 were not always related to strictly scientific interests, but to cross curricular subjects that better reflected the interests of the general public. For example, some of the scientific articles arousing the greatest interest in social networks in 2012 were related to very topical issues such as the Fukushima nuclear accident; cross curricular subjects, such as the effect of coffee consumption on health; or interests closely linked to the profile of a social network user, such as an analysis of classic Nintendo games (Noorden, 2012). Therefore, it is not strange that altmetrics are starting to equate with the social impact of research.

**5. By way of conclusion: current problems for altmetrics.**

Without doubt, altmetrics offers a different outlook when it comes to measuring the visibility, in the widest sense, of scientific and academic papers. These new indicators should be welcomed as being complementary to traditional metrics. However, due to being very new, and only recently applied in scientific contexts, the use of almetrics still has certain limitations that have to be taken into account. Among these is its place within so-called liquid culture, as opposed to solid culture (Area & Ribeiro, 2012). This situation is clearly shown by the evanescent nature of its sources; whereas citation indices such as Web of Science are stable and have trajectories of decades, the same cannot be said of the 2.0 world (Torres-Salinas & Cabezas-Clavijo, 2013). In general, platforms which archive papers, and ultimately generate indicators, usually have very exiguous life cycles and can disappear, as happened with the recent disappearance of Connotea in March 2013. Platforms can also eliminate certain functions, as occurred with Yahoo's removal of the command Search by Site, which shook the foundations of all cibermetrics (Aguillo, 2012). This means that it is currently difficult to choose a reference tool which guarantees medium term continuity. Many uncertainties still exist as to the reproducibility and final significance of results, especially concerning the scientific relevance of the same. This in turn makes it difficult for these tools to be incorporated into the list of evaluative tools.

**Table 5. Examples of the basic characteristics of traditional bibliometric indicators and altmetrics**

| Traditional bibliometric indicators | Altmetrics |
|---|---|
| ● Measure scientific and academic impact through scientific publications, especially articles and journals | ● Measure social impact through means associated with Web 2.0 and not always strictly academic |
| ● Clear association with the concept of scientific recognition and Mertonian normativism | ● Further research is needed to determine the exact significance of the indicators |
| ● Information sources recognised and accepted by the scientific community: Web of Science and Scopus | ● Diverse information sources not always known and used by the whole scientific community |
| ● Sources independently measure the number of citations, subsequently showing various calculations | ● A great variety and heterogeneity of indicators exist, largely dependent on the platforms that produce them |
| ● It is customary to use journal impact indices in order to approximate the quality of scientific articles | ● Indicators highly orientated towards measuring the impact received at article and never journal level |
| ● Bibliometric indicators are highly orientated towards measuring traditional media: articles and books | ● Altmetrics allows measuring of the visibility of less conventional material such as courses or conferences |
| ● Essential referent for agencies and institutions dedicated to the evaluation of scientific activity | ● No agency officially incorporates these methods amongst their indicators for demonstrating the quality of a paper |
| ● Results such as number of citations or a researcher's papers are easily returned in the databases | ● Results are at times difficult to return and are very dependent upon the moment of measurement and the tool |
| ● Measure long term impact, a period of time has to elapse before a publications impact starts to be known | ● Measure immediate impact of a paper in social networks at the moment of publication |
| ● Sometimes they do not function overly well in particular fields, as can be the case of Humanities | ● Can play an important role when it comes to providing measurements in Humanities, where indicators are scarce |

Additionally, the proliferation of sources and users indexing articles aggravates traditional bibliometric problems of normalisation (Haustein & Siebenlist, 2011). In the 2.0 environment, an article can be found indexed or mentioned in multiple ways: by a normalised number, by a URL copied from a web, by part of the title, etc. This causes the compilation of direct mentions, and not indirect article reviews, to be a laborious matter. For example, if an article has been reviewed in a blog, should the diffusion of this entry or its comments be added to the article's original impact? Finally, it has to be mentioned that the empirical study undertaken has also enabled confirmation of the scant concordance of ImpactStory or Almetric.com, which provide different statistics, related only to normalised numbers (DOIs or other type of identifier). Not only is compilation difficult, but also, in most instances, data gathered from many platforms produces very low numbers. Added to this has to be the global difficulty faced by these tools in making data from some of the 2.0 services freely available (Howard, 2012). Despite Adie & Roe (2013) having calculated that more than 2.8 million articles since 2011 have at least one altmetric indicator calculated, the magnitudes provided remain lower than those of citation, even in the majority of cases (see for example the numbers provided in the case studies of Bar-Ilan & al., 2012 or Priem, Piwowar & Hemminger, 2012).

If these indicators are indeed wanted, beyond mere experiments and academic studies, for use in the evaluation of scientific activity, there is no doubt that the many theoretical (significance), methodological (valid sources) and technical (normalisation) problems should still be resolved. These indicators should clearly be used for measuring the social impact of science and, above all, for measuring the impact or immediate visibility of publications, an impossibiity for citation. The new metrics have a very short journey, with an initial burst of activity capturing the visibility of papers at the very moment of publication (Priem & Hemmiger, 2010). This facet complements the classic indicators and even expert reviews, which altmetrics should not aspire to substitute, a situation and a function noted by most scientists (Nature Materials, 2012). Additionally, an identifiable role can be played in fields were bibliometrics is most lacking, as may be the case in humanities (Sula, 2012). It can be stated that new forms of scientific communication require new forms of measurement. For the moment, the only definite conclusion seems to be that altmetrics is here to stay, to enrich the possibilities and dimensions of impact analysis, in all fields of scientific research, and to illuminate from a new perspective the relationship between science and society.